\documentclass{ws-mpla}
\usepackage{lscape}
\begin{document}

\markboth{Kumar, Srivastava, and Morales}
{High-spin structures of $^{77,79,81,83}$As isotopes}
%%%%%%%%%%%%%%%%%%%%% Publisher's Area please ignore %%%%%%%%%%%%%%
\catchline{}{}{}{}{}
%%%%%%%%%%%%%%%%%%%%%%%%%%%%%%%%%%%%%%%%%%%%%%%%%%%%%%%%%%%%%%%%%%%

\title {High-spin structures of $^{77,79,81,83}$As isotopes}

%\title[]{Shell model study of even-even Se isotopes}
\author{VIKAS KUMAR, P.C. SRIVASTAVA\footnote {Emai: pcsrifph@iitr.ac.in}}
\address{Department of Physics, Indian Institute of Technology,
  Roorkee - 247 667, India}
\author{IRVING O. MORALES} 
\address{Instituto de Ciencias Nucleares, Universidad Nacional Aut\'onoma de M\'exico, 04510 M\'exico, D.F., Mexico}
\maketitle

%\pub{Received (Day Month Year)}{Revised (Day Month Year)}

\begin{abstract}
In the present work we report comprehensive set of shell
model calculations for arsenic isotopes.
We performed shell model calculations with two recent effective interactions JUN45 and jj44b.
The overall results
for the energy levels and magnetic moments are in rather good
agreement with the available experimental data. We have also reported competition of
proton- and neutron-pair breakings analysis to identify which nucleon pairs are broken
to obtain the total angular momentum of the calculated states.
Further theoretical development is needed by enlarging model
space by including $\pi 0f_{7/2}$ and $\nu 1d_{5/2}$ orbitals.

\end{abstract}
\ccode{21.60.Cs, 27.50.+e}

%\end{frontmatter}
%=================================================================

%\newpage
\section{Introduction}

The neutron rich nuclei of the Segr\'e chart near Ni region has contributed valuable input for
the understanding of nuclear shell evolution. \cite{nature,pcs,Sorlin,sahin,Sieja2,Franco1,Stefanescu1,Dijon} It is demonstrated that
the shell evolution in nuclei mainly have two types, first type one in which
evolution of nuclear shell as a function of $N$ or $Z$ and second type evolution normally occur within the same nucleus.
\cite{otsuka} Recently experimental evidence for the doubly magic ( $Z=20$ and $N=34$) nature of $^{54}$Ca 
with the onset of a sizable subshell closure at  $N=34$ is reported in Ref. \cite{nature}  
 
%The study of proton excitations across $Z=28$ shell is one of the key issue
%to study evolution of Cu, Ga, As and Br isotopes \cite{lunardi,yoshi}. 
In the case of Cu isotopes it has been shown that around $N=40$ the proton single-particle ordering
changes when neutrons start occupying $g_{9/2}$ orbital. \cite{flangan09}
The systematics of the $1/2^-$, $3/2^-$ and $5/2^-$ levels, magnetic
and electric quadrupole moments between $N=40$ to $N=50$ for Ga isotopes is reported in Ref. \cite{cheal10}
For $^{79}$Ga the $\pi f_{5/2}$ orbital dominant in the g.s. While, for $^{81}$Ga, 5/2$^-$ level become ground state. This is due to
emptying of $\pi p_{3/2}$ orbital to $\pi f_{5/2}$ is started as we move from $^{71}$Ga to $^{79}$Ga.
Ultimately, I$^\pi$=5/2$^-$ become ground state for  $^{81}$Ga. The systematics of low-lying yrast  states
in case of As isotopes is shown in Fig. 1. In case of $^{73-81}$As, the g.s. is $3/2^-$, while for $^{83}$As 
again $5/2^-$ become g.s. In the recent work, Porquet et al
~\cite{astier11} suggested that $f_{5/2}pg_{9/2}$  space is not enough to explain quadrupole excitation built on the 
5/2$_1^-$ and 9/2$_1^+$ state of $^{81}$As. 
Theoretical results for $^{67-79}$As isotopes with projected shell model recently reported  in  
Ref. \cite{preeti13} 

\begin{figure}[h]
\begin{center}
\resizebox{0.8\textwidth}{!}{
\includegraphics{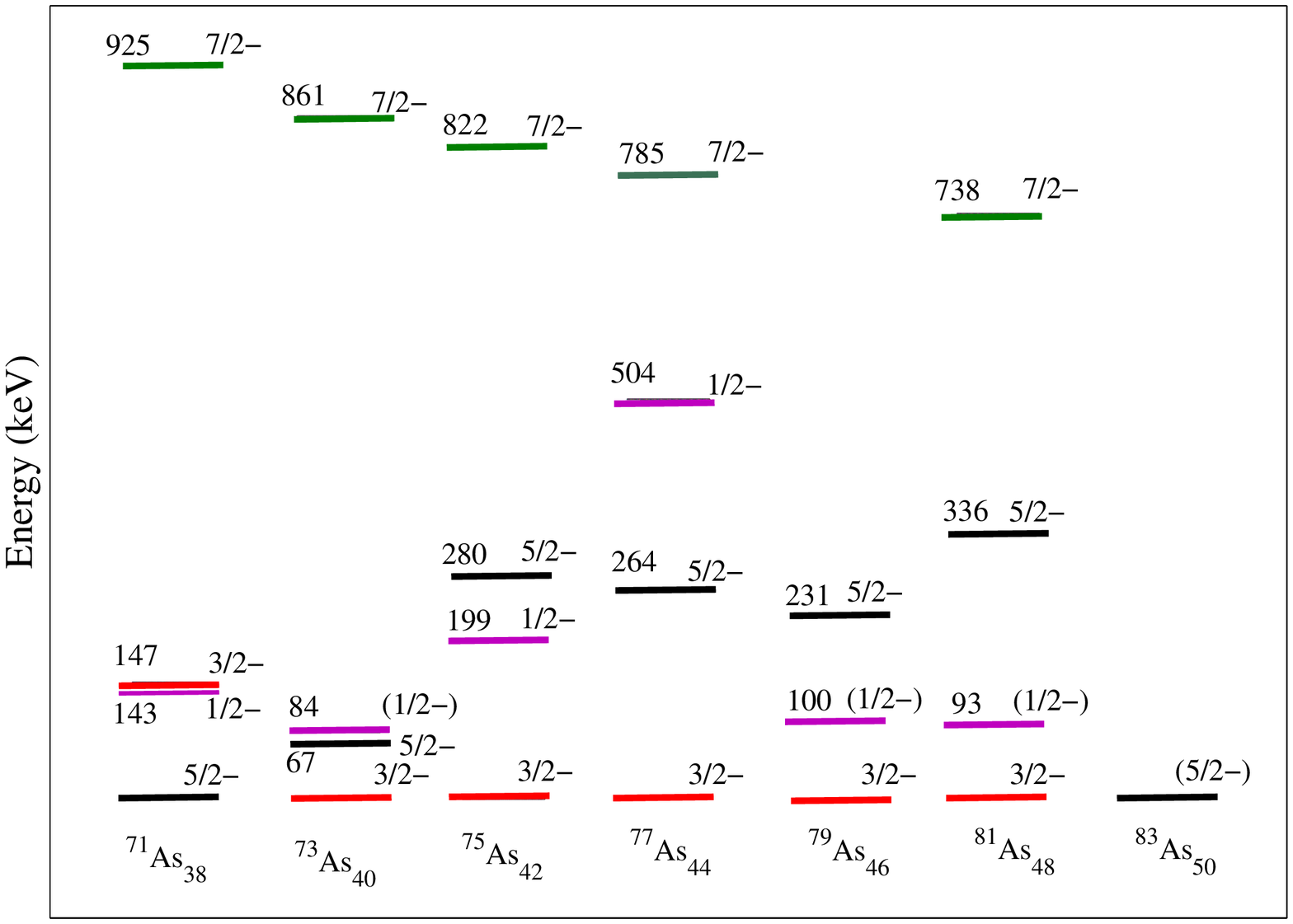} 
}
\caption{\label{f_intro}Low lying yrast states in $^{71}$As to $^{83}$As covering N=40 to N=50 shell closure.
}

\end{center}
\end{figure}

%In section~\ref{details} the details about shell model calculations is described and then spectroscopic
%results for even--even $^{79,81,83,85}$As are presented in section~3. 
%In section 4,  transition probability, quadrupole moments and occupation numbers are presented.
%Finally section~\ref{conc} gives the concluding remarks. 

%%%%%%%%%%%%%%%%%%%%%%%%%%%%%%%%%%%%%%%%%%%%%%%%%%%%%%%%%%%%%%%%%%%%%%%%%%%%%
\section{Outline of Calculations}
\label{details}

In the present work we have performed calculations in the $f_{5/2} \,p \,g_{9/2}$ space.
We have performed calculations with recently available effective interactions
JUN45 \cite{Honma09} and jj44b. \cite{brown} In case of JUN45, the single-particle energies for the 1$p_{3/2}$,
0$f_{5/2}$, 1$p_{1/2}$ and 0$g_{9/2}$ orbits are -9.8280, -8.7087, -7.8388, and -6.2617 MeV
respectively. For jj44b interaction they are -9.6566, -9.2859, -8.2695, and -5.8944
MeV, respectively. The core is $^{56}$Ni, i.e. $N = Z = 28$, and the calculations are performed in
this valence space without truncation. The JUN45 interaction is based on Bonn-C potential, the  single-particle energies and
 two-body matrix elements was modified empirically with A = 63$\sim$69. 
Similarly the jj44b interaction was obtained from a fit to binding energies and excitation
energies with 30 linear combinations of the good $J-T$ two-body matrix elements.
The calculations were performed with shell-model codes ANTOINE \cite{Antoine} and NuShellX.\cite{MSU-NSCL}

%%%%%%%%%%%%%%%%%%%%%%%%%%%%%%%%%%%%%%%%%%%%%%%%%%%%%%%%%%%%%%%%%%%%%%%%%%%%%
%\section{\label{sec3}Spectra}

%\section{Excitation energies for even--even Se isotopes}

\section{Spectra analysis}
The shell model results for $^{77,79,81,83}{\rm As}$ isotopes are presented with respect to the experiment in Figs. 2-5.  

\subsection{$^{77}{\rm As}$}

Comparison of calculated energy levels of $^{77}$As with experimental data is shown in  Fig.~2. The
JUN45 and jj44b interactions predicted  $3/2^-$ level as a ground state which is in good agreement with experiment.
Level $1/2_1^-$ is the first excited state predicted by JUN45 and jj44b. The calculations predicting $1/2_1^-$ level lower
than the experiment. The sequence of the 
experimentally observed energy
levels $3/2_2^-$ and $5/2_1^-$ are same in jj44b while it is interchanged in JUN45. The overall values
of energy levels of $3/2_2^-$ and $5/2_1^-$ with jj44b is in good agreement with experiment. The
calculated $3/2_2^-$ energy level by two effective interactions are higher than experiment. The JUN45 predict
$1/2_2^-$ level at 411 keV, jj44b at 1181 keV, but experiment predicted it at 2195 keV. The
levels $9/2_1^-$ and $7/2_2^-$ calculated by both interactions are in good agreement with experiment. While
levels $5/2_2^-$ and $13/2_1^-$ predicted by two interactions are lower than the experiment. The $7/2_3^-$ level
calculated by JUN45 is in good agreement with experiment while with jj44b it is 303 keV higher
than the experiment.
The sequence of the calculated high-spin negative parity energy levels by JUN45 and jj44b interactions are in
good agreement with experiment.
In the case of positive parity, JUN45 and jj44b predict $9/2_1^+$ as a lowest positive parity state and $5/2_1^+$ as
a second positive parity state which is in good agreement with experiment but these levels are higher than the experimental values.
The level $7/2_1^+$ calculated by JUN45 and jj44b are higher than the
experiment and $7/2_2^+$ level is lower with both JUN45 and jj44b.
The energy separation between $7/2_1^+$ and $7/2_2^+$ is 197 keV in JUN45, 204 keV in jj44b while in the experiment it is 1194 keV. 
Experimentally observed two consecutive levels,
$5/2_2^+$ and $5/2_3^+$ are higher
in both calculations and there are five levels $(7/2_1^+,11/2_1^+,13/2_1^+,7/2_2^+ and 9/2_2^+)$
between them in jj44b.
The JUN45 and jj44b interactions predicted $\pi(p_{3/2}^3)$ configuration for g.s. $3/2_1^-$  with probability 15.7\%
and 13.6\%, respectively.
The structure of negative parity states from $3/2_1^-$ to $19/2_1^-$ is mainly from $\pi(p_{3/2}f_{5/2})^5 $ configuration.
In the case of positive parity, the structure for $9/2_1^+$ state is $\pi(g_{9/2}^1)$, with probability 11.4\% (JUN45)
and 16.9\% (jj44b). The occupancy of $\pi g_{9/2}$ orbital for $9/2_1^+$ state is 1.06 (JUN45) and 1.07 (jj44b).

\subsection{$^{79}{\rm As}$}

In the Fig.~3,  we have shown the comparison of the values of the energy levels calculated by
JUN45 and jj44b with experimental data for $^{77}$As. The JUN45 predict $3/2^-$ level
as a ground state which is in good agreement with experiment, where as jj44b predicts $1/2^-$ as a
ground state which is different from experiment. The levels $3/2_1^-$ and $1/2_1^-$ are interchanged in jj44b.
The levels $3/2_2^-$ and $5/2_1^-$ calculated
by jj44b are in good agreement with experiment values. The calculted $3/2_3^-$ level with jj44b is in good
agreement with experiment while with JUN45 it is lower by 414
keV than the experiment. 

The calculated
value of level $5/2_3^-$ with both interactions are lower than the experiment.
The calculated positive parity levels are higher with both interactions than the
experiment. For positive parity, the JUN45 and jj44b predicts $9/2^+$ as lowest positive parity state which is also predicted by experiment
but it is higher by 656 keV in JUN45 and by 1011 keV in jj44b than the experiment.
As in the experiment the next positive parity state is $5/2^+$ by JUN45, while it is at higher energy
with jj44b interaction. 
In the experiment after $9/2_2^+$ next one is $9/2_3^+$
but in between these two levels JUN45 predict two more levels ($5/2_2^+$ at 2345 keV and $5/2_3^+$ at 2374 keV),
while in jj44b five levels ($1/2_1^+$ at 2129 keV,
$5/2_1^+$ at 2194 keV, $5/2_2^+$ at 2221 keV, $1/2_2^+$ at 2270 keV, $5/2_3^+$ at 2402 keV).
In jj44b calculation the $1/2_1^+$ and $1/2_2^+$ levels are lower than $9/2_3^+$ in comparison
to the experiment value. 

The JUN45 and jj44b interactions predicted  $\pi(p_{3/2}^1)$ configuration for g.s. ($3/2_1^-$) with probability 12.7\%
and 10.1\%, respectively.
The structure of negative parity states from $3/2_1^-$ to $19/2_1^-$ are mainly from $\pi(p_{3/2}f_{5/2})^5 $ configuration.
In the case of positive parity, the structure for $9/2_1^+$ state is $\pi(g_{9/2}^1)$, with probability 16.4\% (JUN45)
and 12.4\% (jj44b). The occupancy of $\pi g_{9/2}$ orbital for $9/2_1^+$ state is 1.07 (JUN45) and 1.06 (jj44b).

\begin{figure*}
\begin{center}
\resizebox{1.2\textwidth}{!}{
\includegraphics{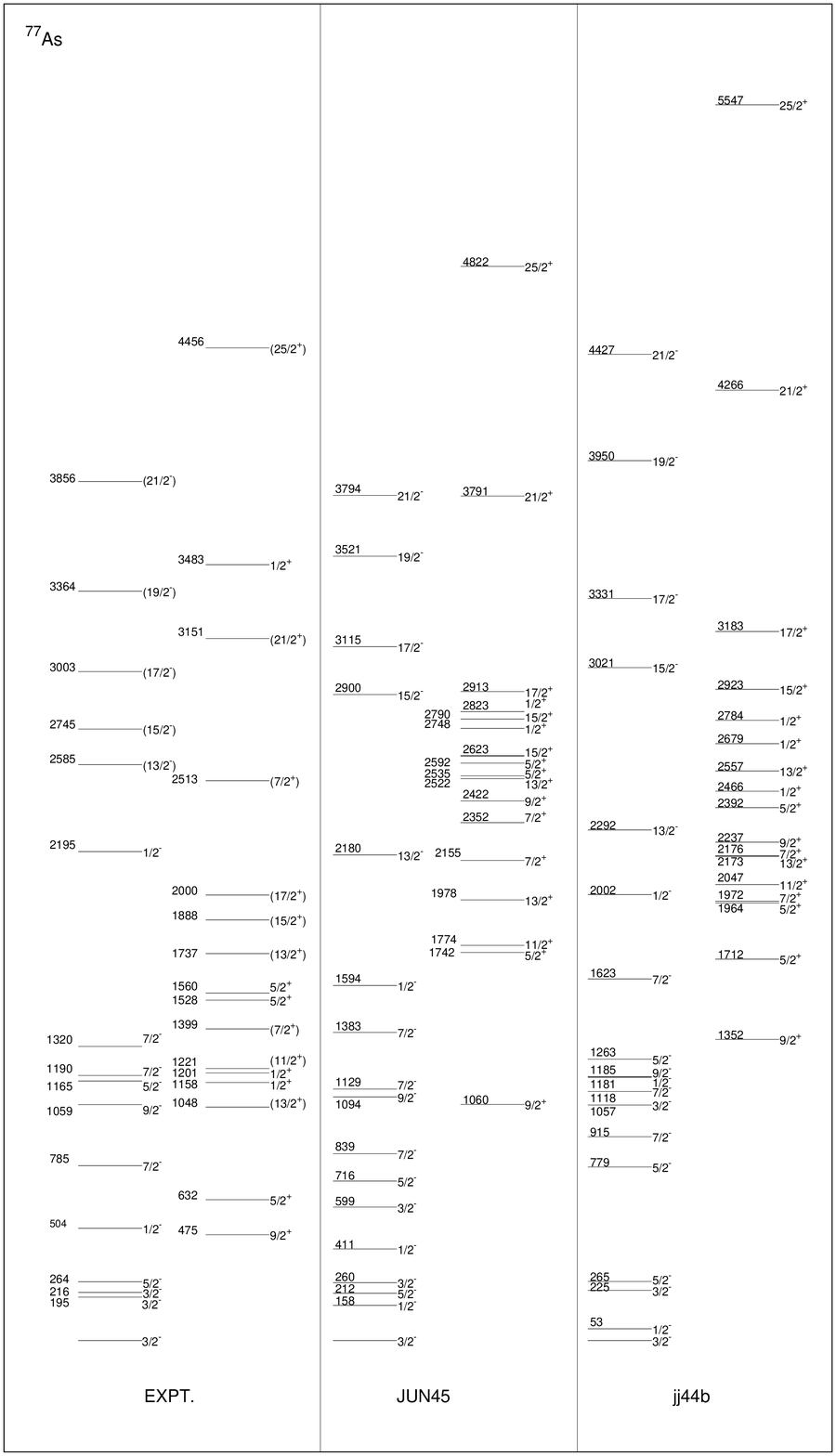} 
}
\caption{\label{f_se78}Comparison of experimental and calculated excitation spectra of $^{77}$As.}

\end{center}
\end{figure*}

\begin{figure*}
\begin{center}
\resizebox{1.2\textwidth}{!}{
\includegraphics{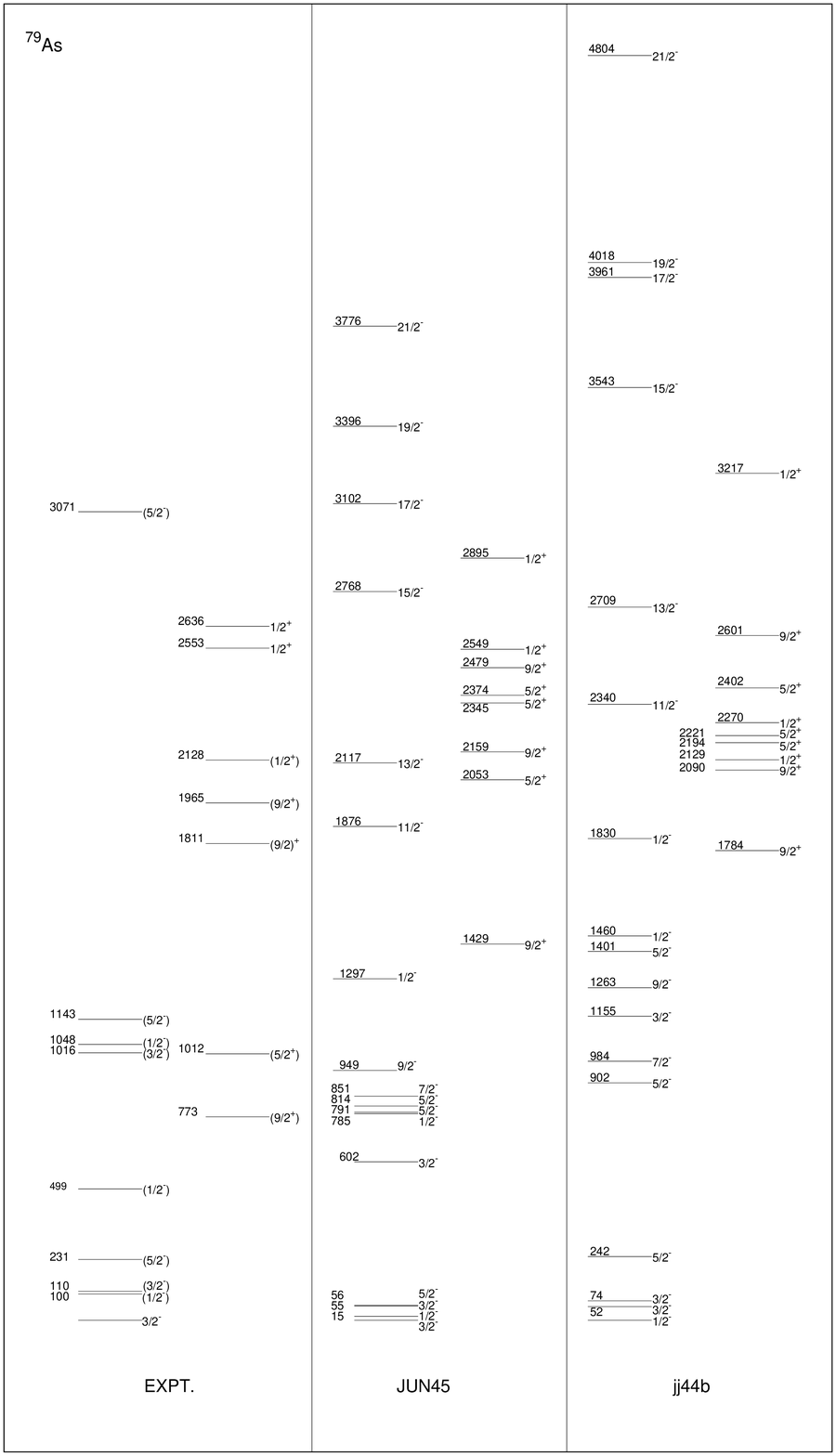} 
}
%\end{center}
\caption{\label{f_se80}Comparison of experimental and calculated excitation spectra of $^{79}$As.}
\end{center}
\end{figure*}

\begin{figure*}
\begin{center}
\resizebox{1.2\textwidth}{!}{
\includegraphics{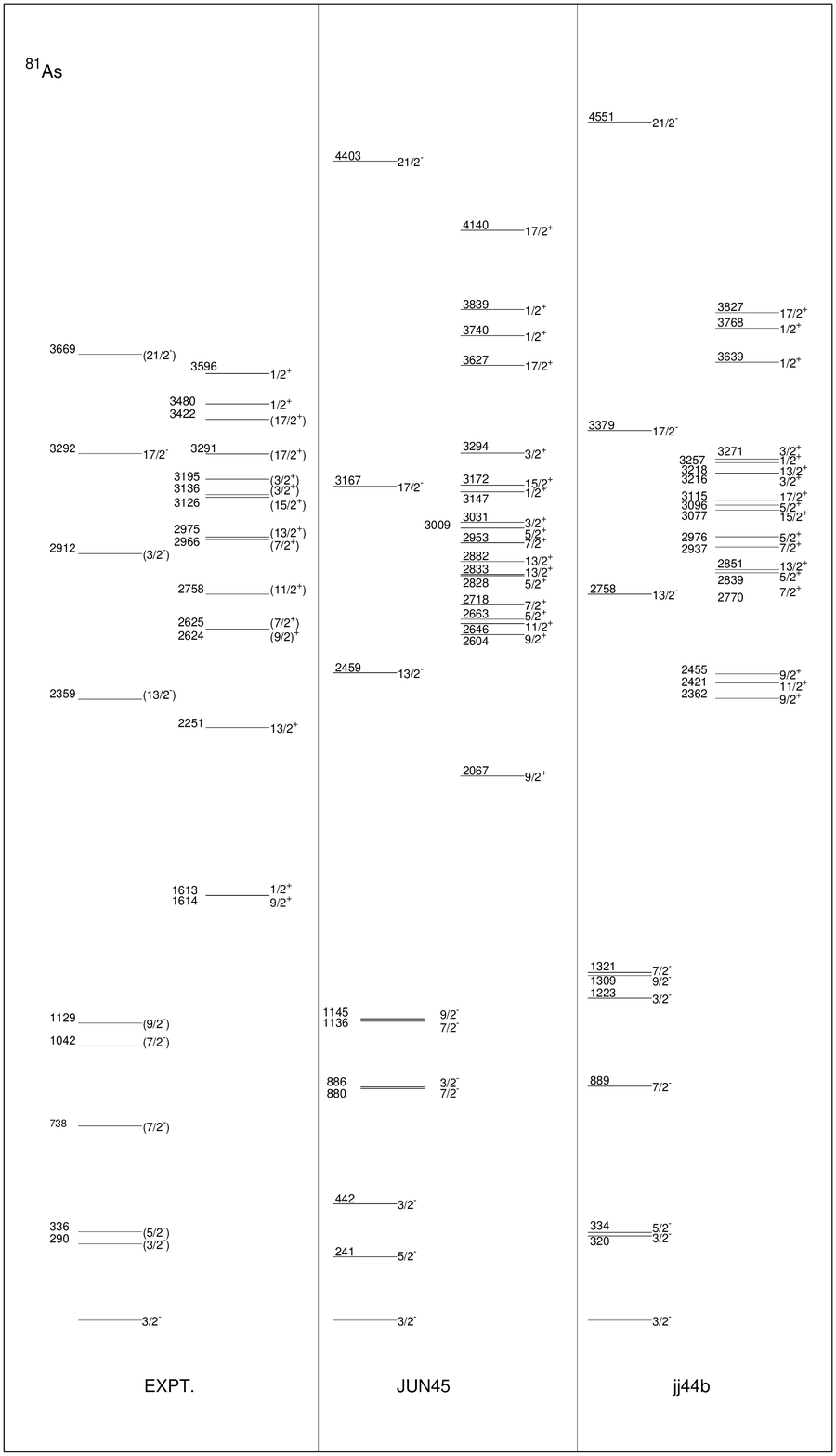} 
}
%\end{center}
\caption{\label{f_se82}Comparison of experimental$^{6}$ and calculated excitation spectra of $^{81}$As.}
\end{center}
\end{figure*}
\begin{figure*}[h]
\begin{center}
\resizebox{1.2\textwidth}{!}{
\includegraphics{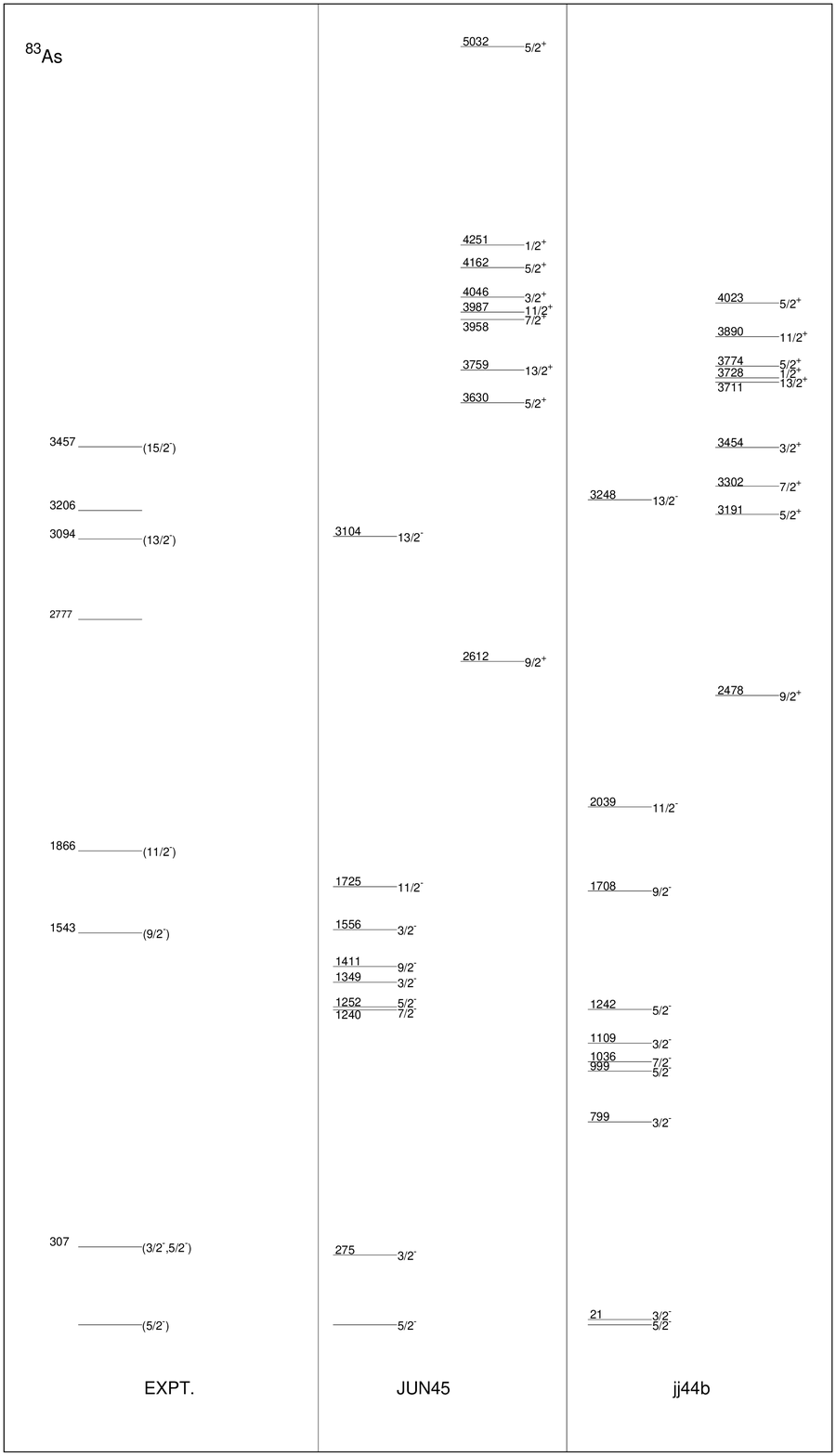} 
}
%\end{center}
\caption{\label{f_se82}Comparison of experimental$^{6}$ and calculated excitation spectra of $^{83}$As.}
\end{center}
\end{figure*}
\subsection{$^{81}{\rm As}$}
 In the Fig. 4~, we have shown results for  $^{81}$As,
the calculated $3/2_2^-$ and $5/2_1^-$ levels by jj44b are in good agreement with experiment, while with JUN45
these levels are interchanged.
The JUN45 and jj44b predicted higher value of $7/2_1^-$ by 142 keV, 151 keV, respectively,
and lower value of $3/2_3^-$ by 2026 keV, 1689 keV  respectively than the experiment. The
level $7/2_2^-$ predicted by JUN45 is higher by 94 keV,  but it is further higher by 279 keV in jj44b. The
calculated $9/2_1^-$ and $13/2_1^-$ levels by JUN45 are in good agreement with experiment but it is predicted
at higher value  by jj44b. The level $17/2_1^-$ calculated by both interactions are in good agreement with experiment.
 The JUN45 and jj44b predicted $21/2_1^-$ level higher by 734 and 882 keV, respectively than the experimental result.

For the positive parity, JUN45 and jj44b predicted  $9/2^+$ as a lowest positive parity state
as in the experiment but this level is at higher energy than the experimental value.
Both the calculations predicting higher value of the level $13/2_1^+$ than the experimental value. 
As we approach towards $N=50$, it is now important to include $\nu 1d_{5/2}$ orbital in the model space
to study neutron excitation across $N=50$ shell.

\subsection{$^{83}{\rm As}$}
We have shown results for $^{83}$As in Fig.~5. Both the interactions 
JUN45 and jj44b predicted g.s. as a $5/2^-$. The levels
$9/2_1^-$ and $11/2_1^-$ are lowered by 132 and 141 keV in JUN45 whereas both levels are
higher by 165 and 173 keV in jj44b. 
With JUN45  $13/2_1^-$ level is in good agreement while in
jj44b it is higher by 154 keV than the experiment.
The sequence of energy levels  $5/2_1^-$, $9/2_1^-$,
$11/2_1^-$ and $13/2_1^-$ in jj44b are in good agreement with experiment. 
The JUN45 and jj44b predicts $9/2^+$ as a lowest positive parity state. While there is no
experimental result is available for positive parity. All the positive parity states predicted by jj44b is lower than JUN45.

%%%%%%%%%%%%%%%%%%%%%%%%%%%%%%%%%%%%%%%%%%%%%%%%%%%%%%%%%%%%%%%%%%%%%%%%%%%%%
\section{ Competition of proton- and neutron-pair breakings analysis}     

In Figs. 6-7 we have shown decomposition of the total angular momentum of the selected states for $^{79,81}$As isotopes with JUN45
interaction. 
From the analysis of the wavefunctions it is possible to identify which nucleon pairs are broken to obtain the total angular
momentum of the calculated states. The two components for neutrons and protons are $I_{n}$ and $I_{p}$ respectively. These
components are coupled to give the total angular momentum of each states. In the Fig. 6(a) -(d), we have shown results of negative parity 
states of $^{79}$As.  The dominant component ( 46 \%) of the 
$3/2_1^-$ ground state comes from $I_{p}$ = 3/2. The $1/2_1^-$ comes from $I_{p}$ = 1/2 (40 \%), $5/2_1^-$ comes from $I_{p}$ = 5/2 (47 \%),
$7/2_1^-$ comes from $I_{p}$ = 7/2 (28 \%). 
The $17/2_1^-$ state has a very peculiar wave function. It shows many different components (35 \% of 5/2$_p^-$ $\otimes$ 6$_{n}^+$, 18 \% of
9/2$_p^-$ $\otimes$ 4$_{n}^+$ ,  14 \% of 3/2$_p^-$ $\otimes$ 8$_{n}^+$ , ...), thus resembling a ''collective'' state. 
The wave function of $19/2_1^-$ state ( predicted at 3396 keV) have the major one corresponding to
$I_{p}$ = 3/2 $\otimes$ $I_{n}$ = 8 ( 44 \%)
and the minor one to $I_{p}$ = 7/2 $\otimes$ $I_{n}$ = 6 ( 14 \%).
Further the wave function of $21/2_1^-$ state ( predicted at 3776 keV) have the major one corresponding to $I_{p}$ = 5/2 $\otimes$ $I_{n}$ = 8 ( 57 \%)
and the minor one to $I_{p}$ = 9/2 $\otimes$ $I_{n}$ = 6 ( 11 \%).
In the Fig. 6(e) -(h), we have shown results of positive parity 
states of $^{79}$As. The major component of $9/2_1^+$ comes from $I_{p}$ = 9/2 (40 \%). Similarly 
$5/2_1^+$ comes from $I_{p}$ = 5/2 (30 \%).

\begin{figure}
\resizebox{160mm}{!}{\includegraphics{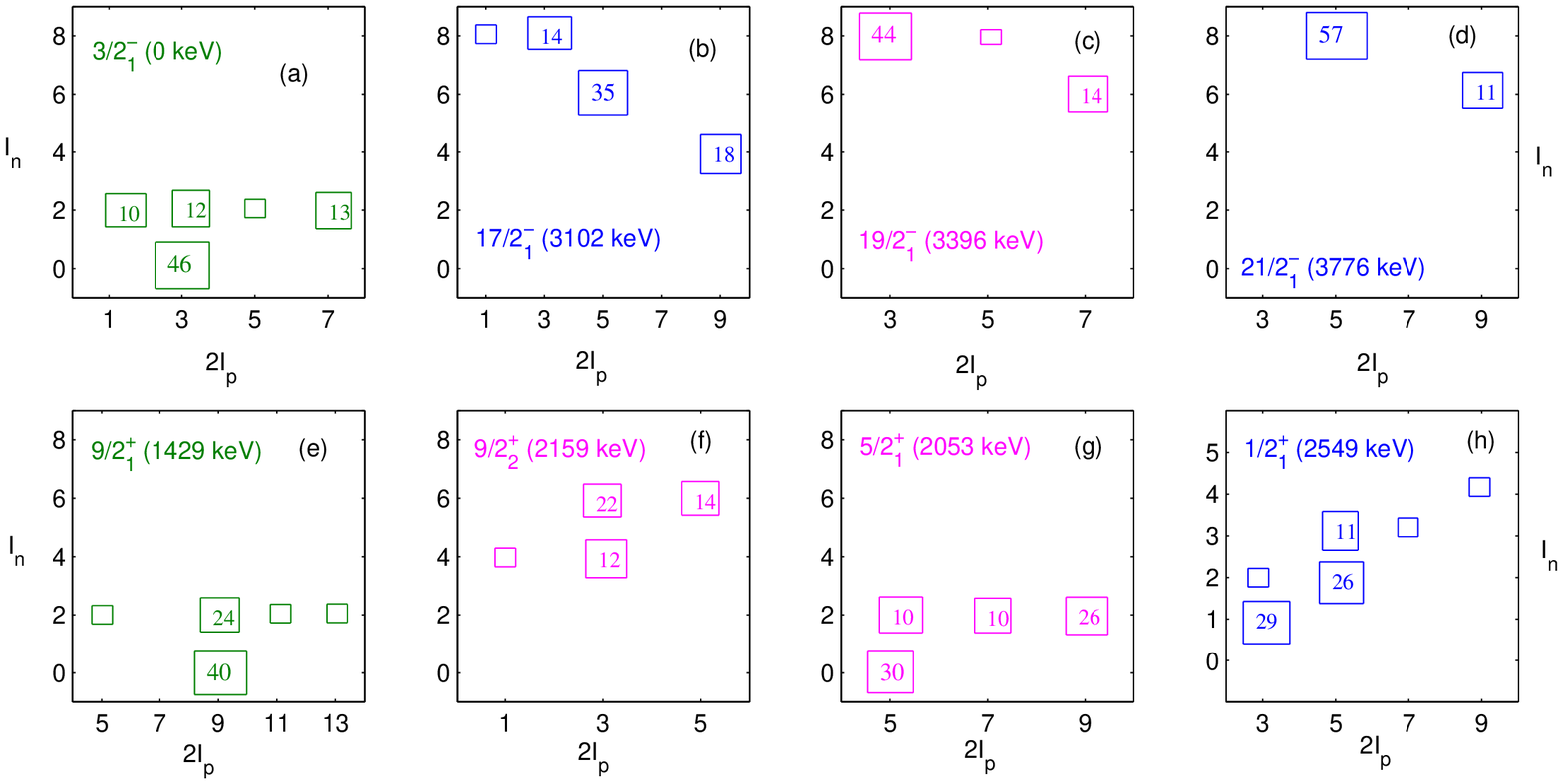}}
\caption{
Decomposition of the total angular momentum of selected states of $^{79}$As  into their $I_n \otimes I_p$ components. The percentage above 10\%
 are written inside the squares, drawn with an area proportional to it. Percentage below 5\% are not written.}
\label{f_79Aspart}
\end{figure}

In the Fig. 7(a) -(d), we have shown results of negative parity 
states of $^{81}$As. The $3/2_1^-$ comes from $I_{p}$ = 3/2 (55 \%). The $17/2_1^-$ comes mainly from $I_{n}$ = 8 ( with 
$I_{p}$ = 1/2 - 7/2 ), i.e. due to neutron pair breaking. Similarly the $21/2_2^-$ ( at 4403 keV) from $I_{n}$ = 8 ( with 
$I_{p}$ = 5/2 - 9/2 ). The $13/2_1^-$ shows many different components (43 \% of 9/2$_p^-$ $\otimes$ 2$_{n}^+$, 23 \% of
5/2$_p^-$ $\otimes$ 4$_{n}^+$ ,  16 \% of 13/2$_p^-$ $\otimes$ 0$_{n}^+$ , ...), thus resembling a ''collective'' state.
The positive parity states results are shown in Fig. 7(e) -(h). The major component of $9/2_1^+$ comes from $I_{p}$ = 9/2 (59 \%).
The $9/2_2^+$ and $17/2_1^+$ is coming from pure neutron breaking. The $1/2_1^+$ shows many different components (65 \%
of 5/2$_p^-$ $\otimes$ 3$_{n}^+$, ...), thus resembling a ''collective'' state.
The above three families are drawn with three different colors, the magenta color is for breaking
of neutron pairs, the green color is for that of protons and blue color is for many components
with various values of $I_{n}$ and $I_{p}$.

\begin{figure}
\resizebox{160mm}{!}{\includegraphics{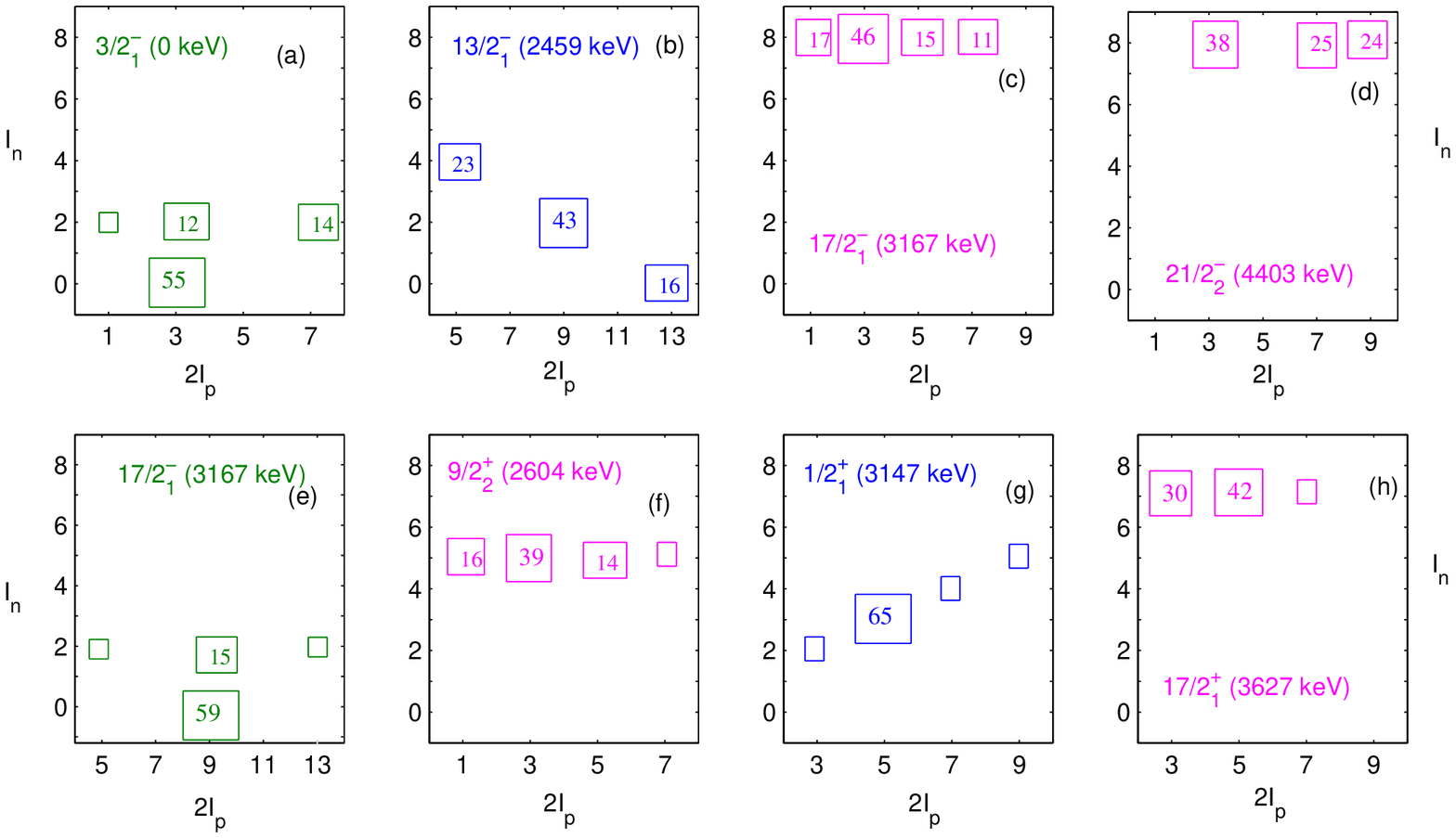}}
\caption{
Decomposition of the total angular momentum of selected states of $^{81}$As  into their $I_n \otimes I_p$ components. The percentage above 10\%
 are written inside the squares, drawn with an area proportional to it. Percentage below 5\% are not written.}
\label{f_81Aspart}
\end{figure}

%%%%%%%%%%%%%%%%%%%%%%%%%%%%%%%%%%%%%%%%%%%%%%%%%%%%%%%%%%%%%%%%%%%%%%%%%%%%%
\section{\label{ep}  Electromagnetic properties and occupation numbers analysis}
\subsection{\label{ep} $E2$ transition probability, quadrupole and magnetic moments}

In Table 1 and 2, we have shown calculated $B(E2)$ and $B(M1)$ values for
different transitions. Although experimental data are very sparse.
For $B(E2$:$5/2_1^+$ $\rightarrow$ $9/2_1^+$), the result of jj44b
interaction is better than JUN45. 
It is further improve by increasing effective charges.
The experimental $B(M1)$ values 
are only available for $^{79}$As, for BM1($5/2_1^- \rightarrow 3/2_1^-$) 
transition the predicted value of JUN45 interaction is close to experimental value.
In Table 3, we have also compared results of quadrupole and
magnetic moments.

%\newpage
The overall results of magnetic moments are in good
agreement with the available experimental data and also the differences between the results obtained by the two adopted
interactions are reasonable. On the other hand it is not possible to draw some
definite conclusions for the $B(E2)$ and quadrupole moments due to lack of
experimental data and contradictory results for the $B(E2)$ values.
The calculated results for $B(E2)$ and quadrupole moments for two interactions are 
different. This may be because two interactions were obtained from a fit with experimental data of different set of nuclei.
The JUN45 interaction  derived by fitting 400 experimental binding
and excitation energy data out of 69 nuclei in the $A =
63 \sim 96$ mass region while jj44b interaction developed by fitting binding
energies and excitation energies from nuclei with $Z = 28 - 30$
and $N = 48 - 50$. The jj44b results are better near $Z = 28$.
The JUN45 interaction is successful along
the $N \sim 50$ isotone chains, while for the Ni region,
the results are not satisfactory, because of the exclusion of
the effect of  $\pi f_{7/2}$ excitations. \cite{Honma09}

%%%%%%%%%%%%%%%%%%%%%%%%%%%%%%%%%%%%%%%%%%%%%%%%%%%%%%%%%%%%%%%%%%%%%%%%%%%%%
\subsection{\label{on} Occupation numbers}

In Fig. 8, we show the proton/neutron occupation numbers.
%Also, the occupancies for the first negative and positive
%parity states are shown in Table 1.
For the ground state ($3/2^-$),
the occupancy of $ \pi (0f_{5/2})$ orbital increase smoothly, while occupancy of 
$\pi(1p_{3/2})$ orbital is decreasing. For $9/2^+$, the occupancy of $\pi (0g_{9/2})$ orbital is significant.
In the case of neutrons orbital the occupancy of the $\nu (0g_{9/2})$ orbital increases  drastically
as neutrons number increase  from $N=44$ ($^{77}$As) 
to $N=50$ ($^{81}$As).

\section{\label{sec5}Summary}
\label{conc}

In summary, comprehensive study for the structure of
neutron-rich odd-even As isotopes have been carried out using large-scale shell-model calculations for
full $f_{5/2}pg_{9/2}$ space with JUN45 and jj44b effective interactions.
The overall results
for the energy levels and magnetic moments are in good
agreement with the available experimental data. The experimental data
for $B(E2)$ and quadrupole moments are not available thus
it is not possible to draw some definite conclusions. 
We have also predicted electromagnetic properties for
other transitions as a guide for future experimental work. 

Further, the following broad conclusions are:

\begin{itemize}
 
\item The results of JUN45 interaction is better than jj44b.

\item High-spin states in $^{77,79,81,83}$As isotopes comes from breaking of neutron/proton pairs.

\item  Further theoretical development is needed by enlarging model
space by including $\pi 0f_{7/2}$ and $\nu 1d_{5/2}$ orbitals to study simultaneously
proton and neutron excitations across $Z=28$  and $Z=50$ shell,
respectively.
\end{itemize} 

\begin{landscape}
\begin{table}
\begin{center}
\title{Table 1. $B(E2)$ reduced transition strength in W.u. Effective charges 
  $e_p=1.5e$, $e_n=0.5e$ / $e_p=1.5e$, $e_n=1.1e$ were used. Experimental values were taken from
  the NNDC database.}
%\vspace{0.2cm}
\label{tab:table2}
%\resizebox{7.8cm}{4.5cm}{
\begin{tabular}{ c | c | c | c | c } \hline %\hline
 & $^{77}$As &$^{79}$As &$^{81}$As & $^{83}$As \\ \hline
BE2($5/2_1^- \rightarrow 3/2_1^-$)  && & &   \\ \hline
Experiment & N/A  & N/A  & N/A  & N/A  \\ %\hline
JUN45 & 5.63 / 9.51 & 1.48 / 2.42 & 0.95 / 1.15 & 0.12 / 0.1818 \\ %\hline
jj44b  &20.92 / 36.90 & 0.06 / 0.11& 1.15 / 1.35& 0.01 / 0.0212 \\ %\hline
%$fpg$  & 9.70 & 12.06 & 0.14 & - \\ %\hline
 && & &  \\ \hline
BE2($3/2_3^- \rightarrow 3/2_1^-$)  && & &   \\ \hline
Experiment & $>$ 4.4  & N/A  & N/A  & N/A  \\ %\hline
JUN45 & 0.42 / 0.66 & 4.19 / 6.67 & 1.56 / 2.62 & 0.89 / 0.8905 \\ %\hline
jj44b  & 0.07 / 0.074 & 1.73 / 3.16 & 1.36 / 2.34 & 4.78 / 4.78 \\ %\hline
%$fpg$  & 0.20 & 0.82 & 0.45 & - \\ %\hline
 && & &  \\ \hline
BE2($7/2_1^- \rightarrow 5/2_1^-$)  && & &   \\ \hline
Experiment & N/A  & N/A  & N/A  & N/A  \\ %\hline
JUN45 & 2.73 / 4.31 & 4.10 / 6.52 & 0.04 / 0.0365 & 2.58 / 2.5839 \\ %\hline
jj44b  &5.78 / 9.75 & 0.94 / 1.48 & 0.09 / 0.1162 & 0.17 / 0.0167 \\ %\hline
%$fpg$  & 3.93 & 0.35 & 0.02 & - \\ %\hline
 && & &  \\ \hline
BE2($9/2_1^- \rightarrow 7/2_1^-$)  && & &   \\ \hline
Experiment & N/A  & N/A  & N/A  & N/A  \\ \hline
JUN45 & 1.62 / 2.52 & 1.72 / 2.39 & 0.40 / 0.4568& 0.52 / 0.5231\\ %\hline
jj44b  & 7.09 /12.33 & 0.11 / 0.18 & 0.08 / 0.0668& 0.58 / 0.5831 \\ \hline
%$fpg$  & 2.87 & 3.44 & 0.52 & -\\ \hline\hline
BE2($5/2_1^+ \rightarrow 9/2_1^+$)  && & &   \\ \hline
Experiment & 80(16) & N/A  & N/A  & N/A  \\ %\hline
JUN45 & 19.39 / 29.78 & 17.79 /27.426 & 12.10 / 18.14 & 5.83 / 5.83\\ %\hline
jj44b  & 23.24 / 38.70 & 7.39 / 11.59 & 12.20 / 18.03& 6.67 /6.67\\ %\hline
%$fpg$  & 1.13 & 3.24 & 2.11 & - \\ %\hline
& & & &  \\ \hline %\hline
BE2($5/2_3^+ \rightarrow 5/2_1^+$)  && & &   \\ \hline
Experiment & $>$ 0.00031 & N/A  & N/A & N/A \\ %\hline
JUN45 & 0.004 / 0.006 & 0.22 / 0.36& 0.21 / 0.30 & 0.46 / 0.46   \\ %\hline
jj44b  & 0.011 /0.021 & 0.26 / 0.49& 0.64  / 1.065& 1.53 / 1.53\\ %\hline
% $fpg$  & 0.64 & 5.04 & 2.26 & - \\ %\hline
& & & &  \\ \hline %\hline
\end{tabular}
\end{center}
\end{table}
\end{landscape}

\begin{landscape}
\begin{table}
\begin{center}
\title{Table 2. $B(M1)$ values for different transitions in W.u. In the present calculation
  $g_s^{eff}$=$g_s^{free}$ / $g_s^{eff}$ = 0.7 $g_s^{free}$ were used. Experimental values were taken from
  the NNDC database.}
%\vspace{0.2cm}
\label{tab:table2}
%\resizebox{7.8cm}{4.5cm}{
\begin{tabular}{ c | c | c | c | c } \hline %\hline
 & $^{77}$As &$^{79}$As &$^{81}$As & $^{83}$As \\ \hline
BM1($3/2_2^- \rightarrow 3/2_1^-$)  && & &   \\ \hline
Experiment & $<$ 0.0041 & N/A  & N/A  & N/A  \\ %\hline
JUN45 & 0.0976 / 0.4782 & 0.0439 / 0.0215 & 0.0414 / 0.0203 & 0.0017 / 0.00083 \\ %\hline
jj44b  &0.0907 / 0.0444 & 0.0021 / 0.0011 & 0.0441 / 0.0216 & 0.3335 / 0.01633 \\ %\hline
%$fpg$  & 0.1765 / & 0.0254 & 0.0022 & - \\ %\hline
 && & &  \\ \hline
BM1($3/2_3^- \rightarrow 3/2_1^-$)  && & &   \\ \hline
Experiment & $>$ 0.0070  & N/A  & N/A  & N/A  \\ %\hline
JUN45 & 0.0606 / 0.0297 & 0.0060 / 0.0029 & 0.0864 / 0.0423& 0.0080 / 0.0079\\ %\hline
jj44b  & 0.0197 / 0.0097 & 0.0027 / 0.0013 & 0.0183 / 0.0090 & 0.0025 / 0.0012 \\ %\hline
%$fpg$  & 0.0333 & 0.0605 & 0.0584 & - \\ %\hline
 && & &  \\ \hline
BM1($5/2_1^- \rightarrow 3/2_1^-$)  && & &   \\ \hline
Experiment & 0.00193(4) & N/A  & N/A  & N/A  \\ %\hline
JUN45 & 0.0016 / 0.00078 & 0.0002 / 0.00011 & 0.0001 / 0.000055 & 0.0002 / 0.00011\\ %\hline
jj44b  &0.0193 / 0.0094 & 0.00005 / 0.0000558 & 0.0006 / 0.000279 & 0.0001 / 0.0000558 \\ %\hline
%$fpg$  & 0.0710 & 0.0209& 0.0002 & - \\ %\hline
 && & &  \\ \hline
BM1($5/2_3^+ \rightarrow 5/2_2^+$)  && & &   \\ \hline
Experiment & $>$ 1.9 x 10$^{-5}$ & N/A  & N/A & N/A \\ %\hline
JUN45 & 0.0660 / 0.0312 & 0.0326 / 0.0159 & 0.000005 / 0.0000 & 0.0012 / 0.0006\\ %\hline
jj44b  & 0.000503 / 0.0002 & 0.1088 / 0.0540 & 0.0227 / 0.0587 & 0.00006 / 0.00005 \\ %\hline
% $fpg$  & 0.02178& 0.0601 & 0.00078 & - \\ %\hline
& & & &  \\ \hline %\hline
\end{tabular}
\end{center}
\end{table}
\end{landscape}

\begin{landscape}
\begin{table}
\begin{center}
\title{Table 3. Comparison of calculated and experimental magnetic and quadrupole moments. The magnetic moments, (in $\mu_N$), with $g_s^{eff}$ = $g_s^{free}$ / $g_s^{eff}$ = 0.7$g_s^{free}$
 and electric quadrupole moments, $Q_s$ (in eb), with $e_p$=1.5e, $e_n$=0.5e / $e_p$=1.5e, $e_n$=1.1e. }
%\vspace{2mm}
\label{tab:table3}
%\resizebox{14.6cm}{4.5cm}{
\begin{tabular}{ c | c | c | c | c } \hline
&$^{77}$As &$^{79}$As &$^{81}$As & $^{83}$As\\ \hline
  $\mu$ ($3/2_1^-$)   && &  & \\ \hline
 Experiment & +1.2946(13) & N/A  & N/A   & N/A \\ %\hline
   JUN45 & +1.941 / +1.358 & +2.825 / +1.977 & +3.090 / +2.167  &+3.666 / +2.566\\ %\hline
   jj44b & +1.477 / +1.034 & +2.668 /+1.868 & +2.985 / +2.089  & +3.569 / +2.498\\ \hline
%$fpg$  & +0.33 & -0.30 & -0.26  & - & $fpg$  & +0.846 & +2.809  & +3.17  & - \\ %\hline
  $\mu$($5/2_1^-$)  && &  & \\ \hline
 Experiment & +0.736 (22)  & N/A   & N/A   & N/A \\ %\hline
  JUN45 & +0.473 / 0.331 & +0.580 / +0.406 & +0.788 / +0.552 & +0.915 / +0.641\\ %\hline
 jj44b & +0.406 / +0.284 & +0.422 /+0.296  & +0.541 /+0.379  & +0.875 / +0.612 \\ \hline
%$fpg$  & +0.45 & +0.05 & +0.36 & -  & $fpg$  & +0.638 & +0.873 & +0.709  & - \\ %\hline
 $\mu$($9/2_1^+$)  & & &  & \\ \hline
 Experiment & +5.525(9)& N/A  & N/A   & N/A \\ %\hline
JUN45 & +5.953 / +4.167 & +6.076 / +4.252 & +6.335 / +4.435 & +6.69 / +4.683\\ %\hline
  jj44b & +5.945 / + 4.1617 & +6.032 / +4.222 & +5.976 / +4.183 & +6.53 / +4.571 \\ %\hline
%$fpg$  & -0.19 & +0.62 & +0.64  & - & $fpg$  & -0.765& +1.941 & -1.346  & -   \\ %\hline
   & & & &\\ \hline \hline
&$^{77}$As &$^{79}$As &$^{81}$As & $^{83}$As       \\  \hline
Q($3/2_1^-$)   && &  &      \\    \hline
Experiment & N/A  & N/A  & N/A   & N/A  \\ 
JUN45 & +0.26 / +0.32 & -0.16 / -0.21 & -0.21 /-0.25  & -0.12 /-0.12\\ 
jj44b & +0.27 / +0.35 & -0.06 / -0.08 & -0.19 / -0.23 & -0.11 /-0.11\\ \hline
Q($5/2_1^-$)  && &  &\\ \hline
Experiment & $<$ 0.75 & N/A   & N/A   & N/A \\ \hline
JUN45 & -0.05 / -0.06 & -0.05 / -0.07 & -0.01 / -0.03  & +0.13 / +0.13\\ 
jj44b & -0.09 / -0.11 & -0.10 / -0.13 & -0.10 /-0.14  & +0.14 /+0.14  \\ \hline
Q($9/2_1^+$)  & & &  & \\ \hline
Experiment & N/A & N/A  & N/A   & N/A  \\ 
JUN45 & -0.74 / -0.97 & -0.68 / -0.83  & -0.538 / -0.612 & -0.387 / -0.386  \\ 
jj44b &-0.75 / -0.97  &  -0.71/ -0.87  & -0.536 / -0.603  & -0.447 / -0.446 \\   
   & & & &\\ \hline 

\end{tabular}
\end{center}
\end{table}
\end{landscape}

\begin{figure}[h]
\begin{center}
\resizebox{120mm}{!}{\includegraphics{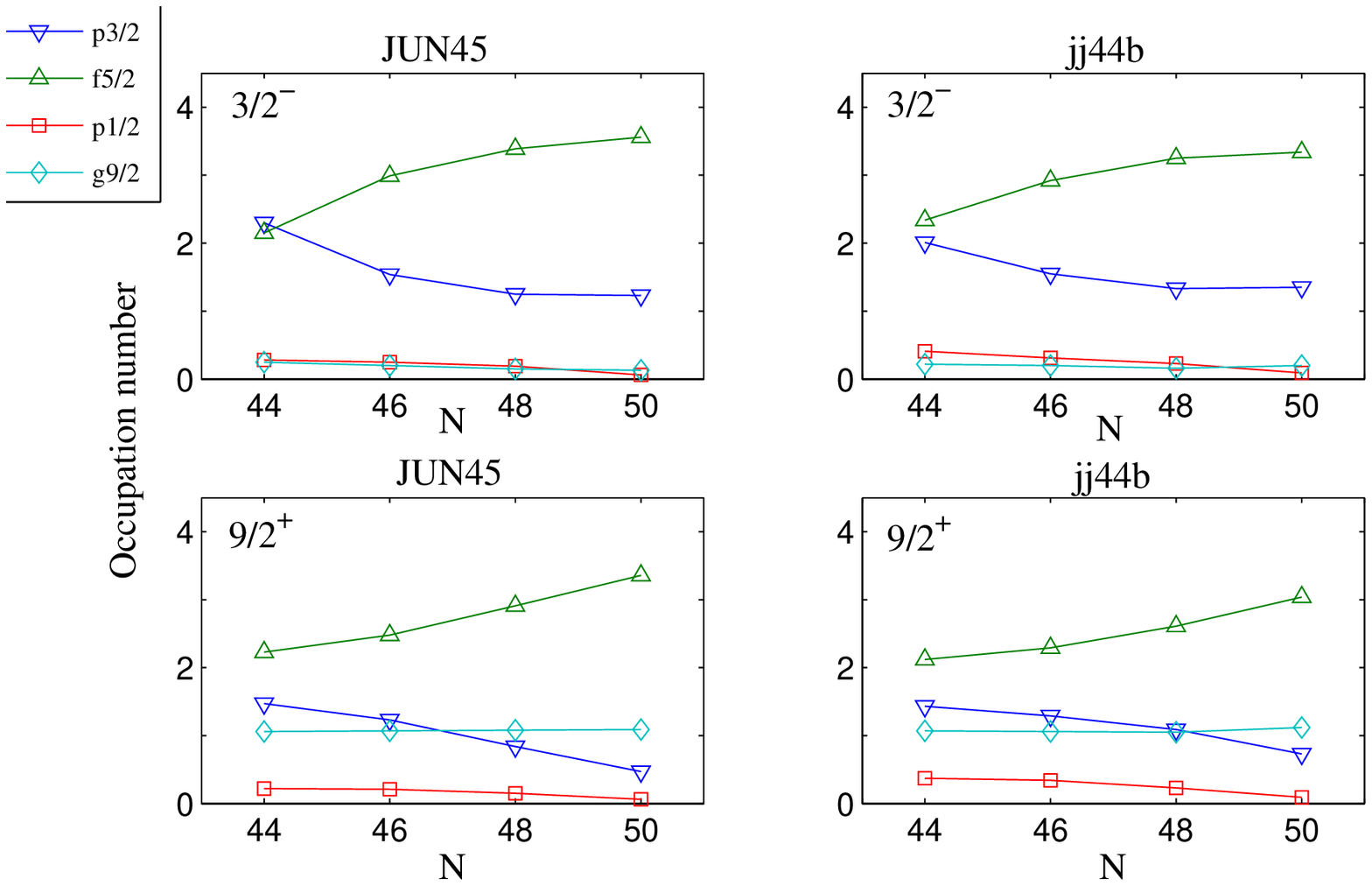}}
\begin{center}
(a) Proton 
\end{center}
\resizebox{120mm}{!}{\includegraphics{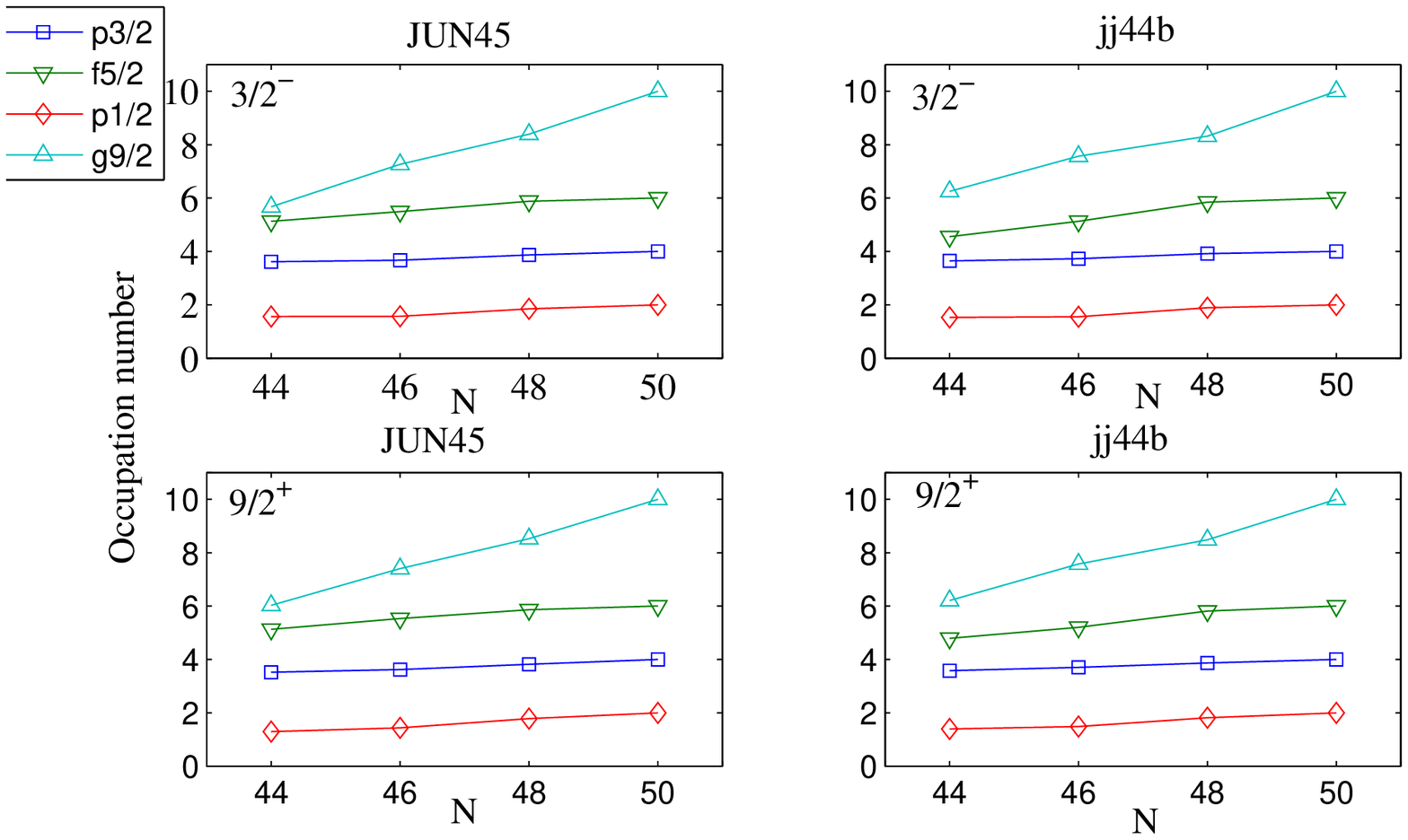}}
\begin{center}
(b) Neutron 
\end{center}
%\epsfig{file=78Ge.eps, angle=-90, width=0.6\linewidth}
%\resizebox{90mm}{!}{\includegraphics{78Ge_1.eps}}
\caption{\label{Fig10}(Color online) Proton/Neutron occupation numbers of the JUN45 and jj44b
($p_{3/2}$, $f_{5/2}$, $p_{1/2}$ and $g_{9/2}$ -shell orbits ) interactions -
 for two low-lying states in $^{77,79,81,83}$As isotopes. Upper
panel for $3/2_1^-$; lower panel for $9/2_1^+$.
}
\end{center}
\label{f_82ge}
\end{figure}
%%%%%%%%%%%%%%%%%%%%%%%%%%%%%%%%%%%%%%%%%%%%%%%%%%%%%%%%%%
{\bf Acknowledgment}

 VK acknowledges partial financial support from CSIR, India for his PhD thesis work.

%{\bf \large  Acknowledgement}
%{\bf \large  References}

\bibliographystyle{elsarticle-num} 

\begin{thebibliography}{00}

\bibitem{nature}
 D. Steppenbeck, {\it et al.}, 
Nature {\bf502} (2013) 207.

\bibitem{pcs}
P.C. Srivastava,  
Mod. Phys. Lett. A {\bf27} (2012) 1250061.

\bibitem{Sorlin}
O. Sorlin and M. -G Porquet,
Prog.\ Part.\ Nucl. \ Phys. {\bf61} (2008) 602. 

\bibitem{sahin}
E. Sahin {\it et al.},
Phys.\ Rev.\ C {\bf91} (2015) 034302. 



\bibitem{Sieja2}
K. Sieja and F. Nowacki,
Phys.\ Rev.\ C {\bf85} (2012) 051301(R).

\bibitem{Franco1}
S. Franchoo {\it et al.},
Phys.\ Rev.\ Lett. {\bf81} (1998) 3100.


\bibitem{Stefanescu1}
I. Stefanescu {\it et al.},
Phys.\ Rev.\ Lett. {\bf100} (2008) 112502.

\bibitem{Dijon}
A. Dijon {\it et al.},
Phys.\ Rev.\ C {\bf85} (2012) 031301.

\bibitem{otsuka}
T. Otsuka, talk at 
FUSTIPEN topical meeting, GANIL, Caen, France, March 17-21, 2014.


\bibitem{flangan09}
 K. T. Flanagan, {\it et al.},
Phys.\ Rev.\ Lett. {\bf103} (2009) 142501. 


\bibitem{cheal10}
B. Cheal, {\it et al.},
Phys.\ Rev.\ Lett. {\bf104} (2010) 252502.

\bibitem{astier11}
M. -G Porquet, {\it et al.},
Phys.\ Rev.\ C {\bf84} ( 2011) 054305.

%\bibitem{kot14}
%J. Kotila and S.M. Lenzi,
%Phys.\ Rev.\ C {\bf89} ( 2014) 064304.


\bibitem{preeti13}
P. Verma, {\it et al.},
Nucl.\ Phys.\ A {\bf918} (2013) 1.

%\bibitem{pcs_ga} 
%P. C. Srivastava,  
%J.\ Phys.\ G\ {\bf39} (2012)  015102. 


%\bibitem{pcs_ge} 
%J. G. Hirsch and P. C. Srivastava,  
%J.\ Phys.\ Conf.\ Ser.  {\bf387} (2012)  012020. 

\bibitem{Honma09}
M. Honma, T. Otsuka , T. Mizusaki  and M. Hjorth-Jensen,  
Phys.\ Rev.\ C {\bf80} (2009) 064323.

\bibitem{brown}
B.A. Brown and A.F. Lisetskiy (unpublished).


\bibitem{Antoine}
E. Caurier,  G. Mart\'inez-Pinedo , F. Nowacki, A. Poves, and A. P. Zuker, 
Rev.\ Mod.\ Phys. {\bf77} (2005) 427. 

\bibitem{MSU-NSCL}
B. A. Brown,  W. D. M. Rae, E. McDonald, M. Horoi, 
NuShellX@MSU.
%\bibitem{kumbartzki14}
%G. J. Kumbartzki, {\it et al.},
%Phys.\ Rev.\ C {\bf89}, ( 2014) 064305.

%\bibitem{kumbartzki12}
%G. J. Kumbartzki, {\it et al.},
%Phys.\ Rev.\ C {\bf85}, ( 2012) 044322.


\end{thebibliography}

\end{document}